% ...
\documentstyle[preprint,revtex,eqsecnum]{aps}
\def\today{\ifcase\month\or
           January\or February\or March\or April\or May\or June\or
           July\or August\or September\or October\or November\or
           December\fi
           \space\number\day, \number\year}
\begin{document}
\draft
\preprint{U. of MD PP \#94-073}
\preprint{DOE/ER/40762-023}
\preprint{\today}
\begin{title}
Rapidity Gap in Jet Events at LEP200
\end{title}

\author{Hung Jung Lu}
\begin{instit}
Department of Physics, University of Maryland\\
College Park, Maryland 20742\\
\end{instit}

\begin{abstract}
We analyze various perturbative mechanisms for the
production of jet events containing rapidity gaps
in $e^+e^-$ annihilation at LEP200
energies. We found that the processes
$e^+e^- \to \gamma^* \gamma^*, \gamma^* Z, Z Z, WW \to 4~ jets$
generate gap events at an observable rate. We point out that LEP200
offers a unique opportunity for the study of gap events
due to a smaller background of ``fake-gaps".
\end{abstract}
\newpage
\narrowtext
\section{Introduction}

High-energy collisions involving hadronic
final states typically create a large number of
particles. A convenient kinematic
quantity for the analysis of many-particle
events is the rapidity variable
\begin{equation}
y=\frac{1}{2}
\ln
\left(  \frac{E+P_\|}{E-P_\|}
\right),
\end{equation}
where $E$ is the energy of a final-state particle
and $P_\|$ its longitudinal momentum with respect
to a reference axis. Rapidity is a natural
longitudinal phase space variable.
(See for instance
\cite{Shapiro,BjorkenPlumber}.) In fact, in
$e^+e^-$ annihilation into jets,
one observes a roughly uniform distribution of
hadrons per unit rapidity as measured with respect
to the jet thrust axis, up to a maximum
value $y_{\rm max} \sim \ln (E_{\rm cm}/m_\pi)$,
where $E_{\rm cm}$ is the total center-of-mass energy.
A recent review on the use of the rapidity variable
as a kinematic analysis tool can be found in Ref.
\cite{BjorkenPlumber,BjorkenGapsJets}.

In hadron-hadron collisions, the distribution of
particles in the rapidity variable often
contains gaps. That is, absence of particles
in a rapidity interval. The existence of a gap
may reflect the underlying exchange mechanism.
For instance, a gap in diffractive scattering
experiments can be interpreted as the
exchange of a pomeron \cite{BjorkenFAD}. The possibility
of using rapidity gaps as triggering signals
in high-mass scale physics has motivated recent
interest in the study of jet events containing
gaps \cite{BjorkenGapsJets,BjorkenFAD,Khoze}.

In hadron-hadron collisions, the analysis of gap
physics is complicated by the presence of
spectator partons. These partons
can interact with each other during the collision,
creating additional particles which may contaminate
and spoil a potential rapidity gap. The survival
probability \cite{BjorkenGapsJets}
of a rapidity gap is an important
subject of study for unraveling of high-mass
scale physics.

In comparison, lepton-initiated collisions offer a
cleaner environment, free of the uncertainties
of survival probability considerations. The study
of rapidity-gap events in lepton colliders can thus offer
important complementary information to the study
of gap events in hadron-hadron machines.

Various perturbative mechanisms
for generating events containing large rapidity
gaps in $e^+e^-$ annihilation
have been analyzed at low energies and
at the $Z$-peak \cite{Randa,BjorkenBrodskyLu,LuBrodskyKhoze}.
Theoretical calculations indicate that these mechanisms
generate gap events at an observable rate
at the $Z$-peak.
The perturbative QCD mechanisms are shown in Fig.
\ref{QCDMechanism}.
Two color-singlet parton-pairs are produced in the final state.
As these parton-pairs move away from each other, the color exchange
between the two dipole systems is suppressed, and hadronization
is dominantly confined to the phase space near each jet-pair.
Hence we expect few or no particles to be produced in the
rapidity interval between the two jet-pair systems.
An interesting QED mechanism is shown in Fig.
\ref{QEDMechanism},
where a small-invariant mass photon decays into
a high-rapidity hadron system, separated by a large
rapidity gap from the other jet-pair. Although this
QED mechanism is suppressed due to the smallness of
$\alpha_{\rm em}$,
its contribution can become important at low photon invariant-mass.
Among all the QCD and QED mechanisms in
Fig. \ref{QCDMechanism} and Fig. \ref{QEDMechanism}, the QCD
$q\bar q q\bar q$ mechanism of Fig. \ref{QCDMechanism} (a)
is the dominant
channel for producing gap events at the $Z$-peak.
As opposed to typical two-jet events in $e^+e^-$ annihilation
which are distributed as $1+\cos^2\theta$ in the scattering
angle of the thrust axis, the jet-pairs produced by the QCD mechanisms
in Fig. \ref{QCDMechanism}
follow a distinctive $\sin^2\theta$ distribution.

Random fluctuations in the final hadron distribution can produce
``fake gaps", which need to be distinguished from the gap
events generated by the partonic mechanisms. The undergoing
Monte Carlo studies \cite{MonteCarlo} are an important
ingredient in the analysis of gap events.

We encounter a unique experimental environment for
studying gap events at LEP200, the future upgrade
of the present $e^+e^-$ collider at CERN.
The four-jet events from the decay of
$\gamma^*\gamma^*, \gamma^*Z, Z Z$ and $WW$
boson pairs occur at a rate comparable to the
two-jet events from of the decay of single $\gamma^*$ and
$Z$ bosons. Hence the ``fake gap" background events
at LEP200 are expected to be reduced
with respect to the situation at the $Z$-peak, where
two- and three-jet events dominate over four-jet events.

The study of gap events at LEP200 will also help
to elucidate the nature of color flow in
quark fragmentation. For the decays of $WW$
into four jets, Gustafson, Peterson and Zerwas
\cite{GustafsonPetterssonZerwas}
have proposed two possible scenarios. In the first
scenario, the color flow is confined to
the quark-antiquark pairs of each gauge boson, and
the two jet-pairs fragment independently.
In the second scenario the string fragmentation
occurs between the quark and antiquarks from
opposite gauge bosons. A measurement of gap event
cross section will therefore constitute an interesting
test of the underlying fragmentation mechanism.

In this paper we study the
$e^+e^- \to \gamma^*\gamma^*, \gamma^*Z, Z Z , WW
\to 4~ jets$ mechanisms for producing gap events
at LEP200 energies. Due to the very different kinematic dependence of
these channels (different flavor combinations,
different invariant-mass resonant regions), we
can neglect interference effects to first approximation,
and consider the four mechanisms separately.
(See also Ref. \cite{FadinKhozeMartin} for a discussion
of interference effects in the production of
heavy unstable particles.)
The previously studied $Z$-peak
mechanisms of Fig. \ref{QCDMechanism} and
Fig. \ref{QEDMechanism} are suppressed
by various kinematic factors (smaller coupling
constant or large off-shellness in the gauge boson propagators),
and do not have observable effects at LEP200.
Our calculations indicate that
the dominant channels for producing gap
events at LEP200 come from $\gamma^*\gamma^*$ and
$\gamma^*Z$ mechanisms. The
$WW$ mechanism has a smaller cross section,
whereas the $Z Z$ mechanism has a negligible
contribution.

\newpage
\section{$\gamma^*\gamma^*$ CASE}

The relevant diagrams are shown in
Fig. \ref{GammaGammaMechanism}. The final
color-singlet jet-pairs $(q_1\bar q_1)~(q_2\bar q_2)$
are effectively the decay
products of the virtual photons. The dominant contribution
of these amplitudes comes from the phase space regions
where the photon propagators acquire large values, that is,
near the photon poles. In fact,
in the small photon invariant-mass limit, the
cross section can be interpreted
as the product of $\sigma(e^+e^- \to \gamma \gamma)$
with the splitting functions of the two photons into respective
$q\bar q$ pairs. Actually, color-singlet jet-pairs
can also be formed by taking the quark from
a virtual photon and the antiquark from the
other virtual photon. That is, we can also form
$(q_1\bar q_2)~(\bar q_1 q_2)$ color-singlet pairs.
However, for gap events the
contribution from these ``exchanged pairs" are strongly
suppressed by the high virtualities carried
by the photons, and also suppressed by an
additional color factor $1/9$.
(This is analogous to the situation
discussed in Ref. \cite{LuBrodskyKhoze} for QED
mechanisms.) Hence, we need only to consider the
contributions from the diagrams in
Fig. \ref{GammaGammaMechanism}.
The expression for the differential
cross section in the small photon invariant-mass limit is
\begin{eqnarray}
d \sigma =
&&
\frac{9 \alpha^4_{\rm em}}
     {4 \pi s^2}
Q^2_{q_1}
Q^2_{q_2}
\frac{d M_1^2}
     {M_1^2}
\frac{d M_2^2}
     {M_2^2}
\Delta(s,M_1^2,M_2^2)
\nonumber
\\
&&
d x_1
\left[ x_1^2 + (1-x_1)^2
\right]
d x_2
\left[ x_2^2 + (1-x_2)^2
\right]
\nonumber
\\
&&
d \cos \theta
\left[ \tan^2(\theta/2) + \cot^2(\theta/2)
\right],
\label{DifferentialCrossSectionGammaGamma}
\end{eqnarray}
where $M_1$ is the invariant mass of the $(q_1 \bar q_1)$ pair,
$M_2$ the invariant mass of the $(q_2 \bar q_2)$ pair,
$x_1$ the longitudinal momentum fraction in the first jet-pair
carried by $q_1$, $x_2$ the longitudinal momemtum fraction
in the second jet-pair carried by $q_2$, $\theta$ the
scattering angle of the thrust axis, $\sqrt{s}$ the
total center-of-mass energy, $Q_{q_1}$ and $Q_{q_2}$ the
electric charge of $q_1$ and $q_2$,
$\alpha_{\rm em}= e^2/ 4 \pi$, with
$e=g \sin \theta_{\rm W}$ the
electromagnetic coupling constant, $g$ the weak coupling
constant and $\theta_{\rm W}$ the weak angle, and finally,
the triangular function is defined by
\begin{equation}
\Delta(s,M_1^2,M_2^2)
=
\sqrt{
         s^2+M_1^4+M_2^4
         -2 s M_1^2 -2 s M_2^2 - 2 M_1^2 M_2^2
     }~.
\label{TriangularFunction}
\end{equation}
The differential cross section is sharply peaked in the
forward and backward beam direction. The apparent singularities
at $\theta \to 0,\pi$ in $\cot(\theta/2)$ and $\tan(\theta/2)$
can be traced back to the electron
propagator in the $t$ and $u$ channel. These singularities
are actually cut off by the photon invariant masses $M_1$ and
$M_2$. For instance, for $\theta \to 0$ one can show that
$t \to -s \theta^2/4 - M_1^2 M_2^2 /s $, hence an appropriate
cut-off value of $\theta$ is
\begin{equation}
\theta_{\rm cut}
=
\frac{2 M_1 M_2}
     {s} .
\end{equation}
An analogous analysis for the $u$ channel propagator
leads to the upper limit $\pi - \theta_{\rm cut}$.
Therefore the angular integral should be limited to the
interval $\theta \in [\theta_{\rm cut},\pi - \theta_{\rm cut}]$.
The final result is not very sensitive to
the precise value of $\theta_{\rm cut}$ because near
the singularities the dependence of the angular integral
on $\theta_{\rm cut}$ is logarithmic:
\begin{equation}
d \sigma
\sim
\int_{\theta_{\rm cut}}
\frac{d \theta}
     {\theta}.
\end{equation}
The cross section is also sharply peaked in the
invariant mass variables $M_1^2$ and $M_2^2$.
The apparent singularities in $M_1^2,M_2^2 \to 0$
in Eq. (\ref{DifferentialCrossSectionGammaGamma})
are physically cut off by the mass threshold of the
$q \bar q$ systems. As in Ref. \cite{LuBrodskyKhoze},
we shall use the values $M^2_\rho,M^2_\omega,M^2_\phi,
M^2_{J/\psi}$ and $M^2_{\Upsilon}$ as representative values
for the invariant mass thresholds for the $u\bar u,
d\bar d, s\bar s, c\bar c$ and $b\bar b$ systems.
To obtain the total cross section, we integrate
over all allowed flavor combinations, noting that
allowed flavors depend on
the integration variables $M_1^2$ and $M_2^2$.
For higher values of $M_1^2$ and $M_2^2$
one has more allowed flavors.
The upper bounds of $M_1^2$ and $M_2^2$ are
given by the kinematic constraint
$\Delta(s, M_1^2, M_2^2) > 0$ in
Eq. (\ref{TriangularFunction}) and also by the
constraints requiring the existence of a
rapidity gap.
The integration limits for $x_1$ and $x_2$ are
set by the gap condition, too. (See Ref.
\cite{BjorkenBrodskyLu,LuBrodskyKhoze}.)

In Fig. \ref{GammaGammaLego}
we present the Lego plot \cite{BjorkenGapsJets}
of a typical
$\gamma^*\gamma^*$ event. Notice that
the two jet-pairs usually have a small
invariant mass, generating thus a large rapidity gap
in the central region. As in Ref.
\cite{BjorkenBrodskyLu} and Ref.
\cite{LuBrodskyKhoze}, we define $g$
as the rapidity difference between the
nearest jet centers from opposite jet-pairs.
Since jet fragments typically are scattered over a radius
$\sim 0.7$ in the Lego plot \cite{BjorkenGapsJets}, the
effective gap is expected to be given by
$g_{\rm eff} \sim g -1.4$.
Virtual photons can easily be transformed
into vector mesons like $\rho, \omega, \phi, J/\psi$
and $\Upsilon$. Hence, we also have the special cases
where a vector meson recoils against a two-jet system
or another vector meson. In particular, the maximum
value of rapidity gap is determined by events where
we have two $\rho$ mesons in the final state, that is,
$g=\ln(s/M_\rho^2)$. For a total center-of-mass
energy of $E=\sqrt{s}=180$ [GeV], the maximum allowed gap
is about 11 units of rapidity.
In Fig. \ref{GammaGammaGapEnergyPlot} we present the
integrated cross section of events having
a value of nearest jet-center rapidity separation
larger than $g$, for different energy levels.
We have used $\alpha_{\rm em} \sim 1/128$ in
our calculation.
In Fig. \ref{GammaGammaEnergyGapPlot}
we plot the integrated cross section
as function of energy, for different values of the
gap-cut $g$. Given a total integrated
luminosity of $500$ [pb$^{-1}$]
per experiment expected for LEP200
\cite{Perkins},
the $\gamma^*\gamma^*$ mechanism should therefore
generate gap events at an observable rate.

Actually, the $\gamma^*\gamma^*$ mechanism also
contributes at the $Z$-peak energy. In the small
invariant mass limit, there is no interference effects
between this mechanism and the resonant QCD and QED
mechanisms of
Fig. \ref{QCDMechanism} and Fig. \ref{QEDMechanism}
in the calculation of the total cross sections.
(The interference terms vanish upon integration
of the azimuthal angles $\phi_1$ and $\phi_2$.
See Appendix for the definition of $\phi_1$
and $\phi_2$. A similar cancellation was
discussed in Ref. \cite{LuBrodskyKhoze}.)
Therefore, each one of them can be obtained
separately. In Fig.
\ref{ZPeakGammaGammaCrossSection} we plot the
integrated gap cross section for the QCD, QED and
the $\gamma^*\gamma^*$ mechanisms at the $Z$ peak
($E=91.17$ [GeV].)
We have used the results from Ref.
\cite{BjorkenBrodskyLu,LuBrodskyKhoze}, and imposed a representative
invariant-mass cut of $30$ [GeV] for the QCD and QED jet-pairs.
We have used a $Z$-peak total visible cross section of $31$ [nb].
The strong coupling constant is taken to be
$\alpha_s \sim 0.13$ in the calculations.
As can be seen from the figure,
the $\gamma^*\gamma^*$ mechanism has small contribution in
the low-gap region, but becomes dominant in the
large-gap region. Physically, the gap events generated by
the $\gamma^*\gamma^*$ are distinct because the jet-pairs
tend to be peaked towards the forward and backward directions,
whereas in the cases of the resonant
QCD and QED mechanisms, the scattering angle is distributed
smoothly as $\sin^2\theta$ and $1+\cos^2\theta$.

\section{$\gamma^*Z$ CASE}
The relevant diagrams are shown in Fig.
\ref{GammaZMechanism}.
This mechanism is similar to the $\gamma^*\gamma^*$
case, but with a virtual photon replaced by a $Z$ boson.
As in the $\gamma^*\gamma^*$ case, we shall not consider
$(q_1 \bar q_2)$ and $(\bar q_1 q_2)$ color singlet pairs, since
the contribution of these ``exchanged pairs" to gap events
is suppressed by a large photon
virtuality and also by the color factor. Due to the
small invariant mass of the virtual photon and the
large mass of the $Z$ boson, we expect the decay
fragments of $\gamma^*$ to be distributed in
the large rapidity region on one side
and those of $Z$ in the
lower rapidity region on the opposite side.
In general we will encounter a lopsided
gap as depicted in Fig. \ref{GammaZLego}.
A special case would be
a vector meson recoiling against the two-jet system
from the $Z$ boson. The maximum
value of rapidity gap
$g=\ln[(s-M_Z^2)/M_\rho M_Z]$
is determined by events where
we have a $\rho$ meson recoiling against
two jets that are distributed symmetrically
with respect to the thrust axis.
For a total center-of-mass
energy of $E=\sqrt{s}=180$ [GeV], the maximum allowed
value of $g$ is about 5.8 units of rapidity.

The differential cross section in the limit of small photon
invariant mass is given by

\begin{eqnarray}
d\sigma
=
&&
\frac{9 \alpha_{\rm em}^2 \alpha_{\rm W}^2}
     {8 \pi}
Q_{q_1}^2
\frac{dM_1^2}
     {M_1^2}
\frac{M_2^2 d M_2^2}
     {(M_2^2 - M_Z^2)^2 + (\Gamma_Z M_Z)^2}
\frac{1}
     {(s-M_2^2)^2}
\Delta(s,M_1^2,M_2^2)
\nonumber
\\
&& dx_1 \ dx_2 \ d\cos\theta
\Biggl\{
\left( 1 + M_2^4/s^2
\right)
\left[ x_1^2 + (1-x_1)^2
\right]
\nonumber
\\
&&
\;\;\;
\left[ \;\;
   \left( {Q^{\rm R}_e}^2 {Q^{\rm R}_{q_2}}^2
        + {Q^{\rm L}_e}^2 {Q^{\rm L}_{q_2}}^2
   \right)
   \left( x_2^2 \cot^2(\theta/2)
        + (1-x_2)^2 \tan^2(\theta/2)
   \right)
\right.
\nonumber
\\
&&
\;\;\;
+
\left.
   \left( {Q^{\rm R}_e}^2 {Q^{\rm L}_{q_2}}^2
        + {Q^{\rm L}_e}^2 {Q^{\rm R}_{q_2}}^2
   \right)
   \left( x_2^2 \tan^2(\theta/2)
        + (1-x_2)^2 \cot^2(\theta/2)
   \right)
\;
\right]
\nonumber
\\
&&
   +
   4 {\bf Q}_e^2 {\bf Q}_{q_2}^2
   \frac{M_2^2}
        {s}
   \left[ x_1^2 + (1-x_1)^2
   \right]
   x_2 (1-x_2)
\Biggr\}.
\label{DifferentialCrossSectionGammaZ}
\end{eqnarray}
In the above expression,
$\alpha_{\rm em}= e^2/ 4 \pi=g^2 \sin^2\theta_{\rm W}/4 \pi $
is the electromagnetic coupling constant and
$\alpha_{\rm W}= g^2/ 4 \pi$ the weak coupling constant,
$M_1$ is the invariant mass of the $(q_1 \bar q_1)$ pair,
$M_2$ the invariant mass of the $(q_2 \bar q_2)$ pair,
$M_Z$ and $\Gamma_Z$ the mass and width of $Z$ boson,
$x_1$ the longitudinal momentum fraction of the virtual
photon carried by $q_1$ and
$x_2$ the longitudinal momentum fraction of the $Z$
boson carried by $q_2$ (see Appendix),
$\theta$ the scattering angle of the thrust axis,
$\sqrt{s}$ the total center-of-mass energy,
$Q_{q_1}$ the electric charge of $q_1$,
$Q^{\rm R}_e$, $Q^{\rm L}_e$, $Q^{\rm R}_{q_2}$ and
$Q^{\rm L}_{q_2}$ the weak charges of $e$ and $q_2$,
where for a fermion of isospin $I_f$ and electric
charge $Q_f$ we have
\begin{eqnarray}
{\bf Q}_f
&=&
{ Q^{\rm L}_f \choose
         Q^{\rm R}_f }
=
{\sec \theta_{\rm W} I_f - \sin \theta_{\rm W} \tan \theta_W Q_f
 \choose
 - \sin \theta_{\rm W} \tan \theta_{\rm W} Q_f } ,
\nonumber
\\
{\bf Q}_f^2
&=&
{Q^{\rm L}_f}^2 + {Q^{\rm R}_f}^2 .
\end{eqnarray}
Due to the narrow width of the $Z$ boson, we can make
the approximation
\begin{equation}
\int
\frac{dM_2^2}
     {(M_2^2 - M_Z^2)^2 + (\Gamma_Z M_Z)^2}
=
\frac{\pi}
     {\Gamma_Z M_Z}
\end{equation}
and replace $M_2^2$ by $M_Z^2$ elsewhere in Eq.
(\ref{DifferentialCrossSectionGammaZ}).

The $\theta$ angle dependence in Eq.
(\ref{DifferentialCrossSectionGammaZ})
implies the existence of a forward-backward
asymmetry in the cross section, which is
also clear from the explicit dependence
on the left-handed and right-handed weak
charges.
The singularities at $\theta \to 0,\pi$ in
$\cot(\theta/2)$ and $\tan(\theta/2)$
can be traced back to the electron
propagator in the $t$ and $u$ channel. These singularities
are physically cut off by the photon invariant mass $M_1$.
For $\theta \to 0$ one can show that
$t \to -(s-M_Z^2) \theta^2/4 - M_1^2 M_Z^2 /(s-M_Z^2)$,
therefore an appropriate cut-off angle is
\begin{equation}
\theta_{\rm cut}
\sim
\frac{2 M_1 M_Z}
     {s-M_Z^2}.
\end{equation}
Similarly, by analyzing the $u$ variable,
one can show that the upper limit for
the $\theta$ integration should be $\pi - \theta_{\rm cut}$.
Define
\begin{equation}
I_\theta(M_1^2)
=
\int_{\theta_{\rm cut}}^{\pi - \theta_{\rm cut}}
d \cos \theta \tan^2(\theta/2)
=
\int_{\theta_{\rm cut}}^{\pi - \theta_{\rm cut}}
d \cos \theta \cot^2(\theta/2) ,
\end{equation}
the expression of the cross section in Eq.
(\ref{DifferentialCrossSectionGammaZ})
simplifies to
\begin{eqnarray}
d\sigma
=
&&
\frac{9 \alpha_{\rm em}^2 \alpha_{\rm W}^2}
     {8} \
{\bf Q}_e^2
Q_{q_1}^2
{\bf Q}_{q_2}^2 \
\frac{M_Z}
     {\Gamma_Z(s-M_Z^2)^2}
\nonumber
\\
&&
\frac{dM_1^2}
     {M_1^2} \
\Delta(s,M_1^2,M_Z^2)
\nonumber
\\
&&
dx_1
\left[ x_1^2 + (1-x_1)^2
\right]
\nonumber
\\
&&
dx_2
\left\{
   \left( 1 + M_Z^4/s^2
   \right)
   \left[ x_2^2 + (1-x_2)^2
   \right]
   I_\theta(M_1^2)
\right.
\nonumber
\\
&&
\left.
+ \ 8 \ (M_Z^2/s) \ x_2 (1-x_2)
\right\} .
\label{SimplifiedDifferentialCrossSectionGammaZ}
\end{eqnarray}
To obtain the total cross section, we
sum over all allowed flavor combinations.
As in the $\gamma^*\gamma^*$ case,
we use the values $M^2_\rho,M^2_\omega,M^2_\phi,
M^2_{J/\psi}$ and $M^2_{\Upsilon}$ as representative values
for the invariant mass thresholds for the $u\bar u,
d\bar d, s\bar s, c\bar c$ and $b\bar b$ systems.
The $x_1$, $x_2$ integrations, as well as the $M_1^2$
integration, are also limited by the gap condition
\cite{BjorkenBrodskyLu,LuBrodskyKhoze}.
In Fig. \ref{GammaZGapEnergyPlot} we plot the
integrated cross section of events having
a value of nearest jet-center rapidity separation
larger than $g$, for different energy levels.
We have used $\alpha_{\rm W} \sim 1/29.8$
and $\sin^2\theta_{\rm W}=0.233$ in
our calculation.
In Fig. \ref{GammaZEnergyGapPlot} we plot the
integrated cross section
as function of energy, for different values of the
gap-cut $g$. For the projected integrated
luminosity of $500$ [pb$^{-1}$]
per experiment at LEP200,
the $\gamma^*Z$ mechanism therefore also
generates an observable number of gap events.
For smaller values of $g$ ($g < 4$) the
$\gamma^* Z$ mechanism dominates over the
$\gamma^*\gamma^*$ mechanism. As discussed
previously, the $\gamma^*Z$ mechanism cannot
not generate gap events with $g$ larger than
$\sim 5.8$, hence for large gaps ($g > 5$)
the $\gamma^*\gamma^*$ mechanism dominates.

\newpage
\section{$Z Z$ CASE}

As opposed to the $\gamma^*\gamma^*$ and
$\gamma^* Z$ mechanisms, the direct jet-pairs
from the decay of $ZZ$ and $WW$ particles are not capable
of generating events with large rapidity gaps.
This is because at LEP200 the $ZZ$ and $WW$ pairs
are produced right above the threshold, thus
the two gauge bosons recoil slowly against
each other. For instance, at a center-of-mass energy
of 200 [GeV], we have only 1.4 units of rapidity separating
two recoiling $W$ particles, which means in
practice no observable rapidity gap. However, the
``exchanged pairs" as those shown in Fig. \ref{ZZMechanism}
provide a feasible channel of producing gap events.
It is interesting to recall that the exchanged-pair
contributions are negligible in the $\gamma^*\gamma^*$ and
$\gamma^* Z$ cases due to the large invariant mass carried
by the virtual photons, therefore the direct pairs are
responsible for gap events there. The situation is exactly reversed
for the $ZZ$ and $WW$ cases, where the exchanged pairs
are the responsible mechanism.

Let us consider first the case of a $ZZ$ pair produced exactly
at threshold. In the  center-of-mass frame,
each of these static $Z$ particles decays into a back-to-back
quark-antiquark pair. In the phase space where $q_1$ becomes
collinear with $\bar q_2$ (hence $\bar q_1$ becomes also
collinear with $q_2$), we can form small-invariant-mass
color-singlet jet pairs and have a rapidity gap.
Naturally, to have an observable cross section, the
center-of-mass energy should be larger than the
threshold value. As the center-of-mass
energy becomes higher and higher, the crossing of
a quark from the phase space of
a $Z$ boson to the phase space of the
opposite $Z$ boson becomes less and less likely. Therefore,
we expect the gap event cross section to rise from zero
to a maximum value somewhere above the threshold, and then
to decrease with increasing center-of-mass energy.

The crossing of particles in the kinematic
phase space substantially complicates the calculation
of the gap event cross section. This is because
the gap condition becomes a unnatural cut in the
integration of the various kinematic variables.
In particular, as opposed to the $\gamma^*\gamma^*$
case and the $\gamma^*Z$ case,
the azimuthal angles $\phi_1$ and
$\phi_2$ (see Appendix) can no longer be integrated
out analytically.

We compute the helicity amplitudes in terms of
spinor products defined in Ref.
\cite{ManganoParke}. After using the narrow
width approximation for the $Z$ boson propagators
to integrated out the jet-pair invariant masses
$M_1^2$ and $M_2^2$, we obtain
\begin{eqnarray}
d \sigma
&=&
\frac{\alpha_{\rm W}^4}
     {\pi (32  \Gamma_Z M_Z \ s)^2} \
\Delta(s,M_Z^2,M_Z^2) \
d x_1   d x_2   d \phi_1   d\phi_2 d \cos \theta
\nonumber
\\
& &
\sum_{q_1,q_2}
\sum_{H_e,H_{q_1},H_{q_2}}
{Q_e^{H_e}}^4
{Q_{q_1}^{H_{q_1}}}^2
{Q_{q_2}^{H_{q_2}}}^2
\left|
   T(H_e,H_{q_1},H_{q_2})
\right|^2 ,
\label{DifferentialCrossSectionZZ}
\end{eqnarray}
where the first sum is over all allowed quark flavors,
and the second sum is over all helicities of the
electron and the final quarks $q_1$ and $q_2$.
We have included in the formula an $1/2$ factor
to account for the double counting of flavors
in the summation of $q_1$ and $q_2$. The various
charges in Eq. (\ref{DifferentialCrossSectionZZ})
are the weak charges of the electron, $q_1$ and $q_2$,
as defined in the previous section.
(We use indifferently $(R,L)$ and $(+,-)$
to denote right and left helicity.)
There are eight helicity amplitudes, but half of them
are needed since the other half
can be obtained by simple helicity conjugation:
\begin{eqnarray}
T(+++)
=
&&
\frac{4}
     {t} \
[k_1 q_1] \langle k_2 \bar q_2 \rangle \
\Bigl\{
-[k_1 q_2] \langle k_1 \bar q_1 \rangle +
 [q_1 q_2] \langle q_1 \bar q_1 \rangle
\Bigr\}
\nonumber
\\
+
&&
\frac{4}
     {u} \
[k_1 q_2] \langle k_2 \bar q_1 \rangle \
\Bigl\{
-[k_1 q_1] \langle k_1 \bar q_2 \rangle +
 [q_1 q_2] \langle \bar q_2 q_2 \rangle
\Bigr\} ,
\nonumber
\\
T(++-)
=
&&
\frac{4}
     {t} \
[k_1 q_1] \langle k_2 q_2 \rangle \
\Bigl\{
-[k_1 \bar q_2] \langle k_1 \bar q_1 \rangle +
 [q_1 \bar q_2] \langle q_1 \bar q_1 \rangle
\Bigr\}
\nonumber
\\
-
&&
\frac{4}
     {u} \
[k_1 \bar q_2] \langle k_2 \bar q_1 \rangle \
\Bigl\{
 [k_1 q_1] \langle k_1 q_2 \rangle +
 [q_1 \bar q_2] \langle \bar q_2 q_2 \rangle
\Bigr\} ,
\nonumber
\\
T(+-+)
=
&&
-
\frac{4}
     {t} \
[k_1 \bar q_1] \langle k_2 \bar q_2 \rangle \
\Bigl\{
 [k_1 q_2] \langle k_1 q_1 \rangle +
 [\bar q_1 q_2] \langle q_1 \bar q_1 \rangle
\Bigr\}
\nonumber
\\
+
&&
\frac{4}
     {u} \
[k_1 q_2] \langle k_2 q_1 \rangle \
\Bigl\{
-[k_1 \bar q_1] \langle k_1 \bar q_2 \rangle +
 [\bar q_1 q_2] \langle \bar q_2 q_2 \rangle
\Bigr\} ,
\nonumber
\\
T(+--)
=
&&
\frac{4}
     {t} \
[k_1 \bar q_1] \langle k_2 q_2 \rangle \
\Bigl\{
-[k_1 \bar q_2] \langle k_1 q_1 \rangle +
 [\bar q_2 \bar q_1] \langle q_1 \bar q_1 \rangle
\Bigr\}
\nonumber
\\
+
&&
\frac{4}
     {u} \
[k_1 \bar q_2] \langle k_2 q_1 \rangle \
\Bigl\{
-[k_1 \bar q_1] \langle k_1 q_2 \rangle +
 [\bar q_2 \bar q_1] \langle \bar q_2 q_2 \rangle
\Bigr\} .
\end{eqnarray}
The spinor products in these amplitudes can be
evaluated numerically with the
formulas given in the Appendix.

Due to the crossing of quark pairs, the thrust
axis in general is not aligned with the gauge
boson direction. We will denote the scattering
angle of the thrust axis by $\Theta$, which
in general differs from the gauge boson $\theta$ angle
that appears in the expression of the cross section
of Eq. (\ref{DifferentialCrossSectionZZ}).
In the Monte Carlo integration of the cross section in Eq.
\ref{DifferentialCrossSectionZZ}, we only retain those
events containing a rapidity gap larger than the gap cut
$g$. The result is plotted in Fig.
\ref{ZZEnergyGapPlot}.
We notice that large gap events from the $ZZ$ mechanism
have cross sections much smaller than $0.01$ [pb]. Hence,
this mechanism is not expected to have observable
effects at the projected luminosity of LEP200.
In Fig. \ref{ZZAngularDistributionPlot}
we plot the angular distribution of the cross section
for $E=192$ [GeV] and $g=3$. We see that the cross section
has two smooth humps,
as opposed to the $\gamma^*\gamma^*$ and $\gamma^*Z$
cases where the cross section has two sharp peaks $\sim 1/\theta$
in the forward and backward beam directions.

\section{$W W$ CASE}

The kinematics of the $WW$ case is similar to
the $ZZ$ case. Diagrammatically, the production of
$W W$ boson pairs is interesting because it
involves the triple-gauge-boson vertex
of the electroweak theory. The Feynman diagrams
of this mechanism is shown in Fig.
\ref{WWMechanism}.
The $u_1$, $\bar u_2$, and $\bar d_1$, $d_2$ quarks
represents any up-type $(u,c)$ and down-type $(d,s,b)$
quarks and antiquarks. As explained in the previous
section, at LEP200 the direct-pair contributions
are not capable of producing events with large rapidity
gaps, therefore we need only to focus on the exchanged-pair
contributions of Fig. \ref{WWMechanism}.
Like in the $ZZ$ case, we also expect the gap event
cross section to reach a maximum somewhere above the
threshold, since at higher energies there is less
phase space for the formation of exchanged pairs.

The differential cross section for the $WW$ case is
given by
\begin{eqnarray}
d \sigma
=
&&
\frac{\alpha_{\rm W}^4}
     {512\pi (\Gamma_Z M_Z s)^2} \
\Delta(s,M_W^2,M_W^2) \
d x_1 d x_2 d \phi_1 d \phi_2 d \cos \theta
\nonumber
\\
&&
\Biggl\{ \
\left|T_1^+
\right|^2
\left|
        \frac{1}
             {s}
       -\frac{1}
             {s-M_Z^2 + i \Gamma_Z M_Z}
\right|^2
\sin^4\theta_{\rm W}
\nonumber
\\
&&
+
\left|
   \left(
          \frac{\sin^2\theta_{\rm W}}
               {s}
        + \frac{\frac{1}{2}-\sin^2\theta_{\rm W}}
               {s-M_Z^2 + i \Gamma_Z M_Z}
   \right) T_1^-
         -\frac{1}
               {2 t} T_2^-
\right|^2 \
\Biggr\} ,
\label{DifferentialCrossSectionWW}
\end{eqnarray}
where $M_W$ and $\Gamma_Z$ are the mass and width
of the $W$ boson, $\Delta$ as defined in Eq.
(\ref{TriangularFunction}). We have summed over
all allowed quark flavors and made following
approximation for the summation over the
Cabbibo-Kobayashi-Maskawa matrix elements
(See for instance Ref.
\cite{ReviewParticleProperties}):
\begin{equation}
\sum_{{i=u,c} \atop {j=d,s,b}}
\left| V_{ij}
\right|^2
\sim 2 .
\end{equation}
The $T_1^+$ and $T_1^-$ amplitudes
come from the $s$-channel
diagram in Fig.
\ref{WWMechanism} (a), and are given by
\begin{eqnarray}
T_1^+
&=&
4 \     [\bar d_1 \bar u_2] \langle d_2 u_1 \rangle \
  \Bigl\{
        [k_1 u_1] \langle u_1 k_2 \rangle
      + [k_1 \bar d_1] \langle \bar d_1 k_2 \rangle
  \Bigr\}
\nonumber
\\
&+&
4 \     [k_1 \bar u_2] \langle d_2 k_2 \rangle \
  \Bigl\{
        [\bar d_1 \bar u_2] \langle \bar u_2 u_1 \rangle
      + [\bar d_1 d_2] \langle d_2 u_1 \rangle
  \Bigr\}
\nonumber
\\
&-&
4 \     [k_1 \bar d_1] \langle u_1 k_2 \rangle \
  \Bigl\{
        [\bar u_2 u_1] \langle u_1 d_2 \rangle
      + [\bar u_2 \bar d_1] \langle \bar d_1 d_2 \rangle
  \Bigr\} ,
\nonumber
\\
T_1^-
&=&
4 \     [\bar d_1 \bar u_2] \langle d_2 u_1 \rangle \
  \Bigl\{
        [k_2 u_1] \langle u_1 k_1 \rangle
      + [k_2 \bar d_1] \langle \bar d_1 k_1 \rangle
  \Bigr\}
\nonumber
\\
&+&
4 \     [k_2 \bar u_2] \langle d_2 k_1 \rangle \
  \Bigl\{
        [\bar d_1 \bar u_2] \langle \bar u_2 u_1 \rangle
      + [\bar d_1 d_2] \langle d_2 u_1 \rangle
  \Bigr\}
\nonumber
\\
&-&
4 \     [k_2 \bar d_1] \langle u_1 k_1 \rangle \
  \Bigl\{
        [\bar u_2 u_1] \langle u_1 d_2 \rangle
      + [\bar u_2 \bar d_1] \langle \bar d_1 d_2 \rangle
  \Bigr\} ,
\end{eqnarray}
where we have used the symbol $u_1$, $\bar d_1$,
$\bar u_2$ and $d_2$ to denote the final state quarks
as well as their respective final momenta. The
$T_2^-$ amplitude is given by
\begin{equation}
T_2^-
=
4 \     [u_1 k_1] \langle k_2 \bar u_2 \rangle \
  \Bigl\{
        [u_1 \bar d_1] \langle d_2 u_1 \rangle
      - [k_1 \bar d_1] \langle d_2 k_1 \rangle
  \Bigr\} .
\end{equation}
The spinor products in these expressions can be evaluated
with the formulas given in the Appendix.

As in the $ZZ$ case, the thrust
axis angle $\Theta$ in general differs from the gauge boson angle
$\theta$ in the expression of the cross section
of Eq. (\ref{DifferentialCrossSectionWW}).
In the Monte Carlo integration of the cross section
in Eq. (\ref{DifferentialCrossSectionWW})
we only retain those events with a rapidity gap
larger than the gap cut value $g$. The resulting
cross section is given in Fig. \ref{WWEnergyGapPlot}.
We see from the plot that the $WW$ mechanism
can contribute somewhat to the production of gap
events at LEP200. For gap cuts $g$ between $3$ and $4$
units of rapidity, the maximum cross section is reached
around $5$ to $10$ [GeV] above the threshold.
In Fig. \ref{WWAngularDistributionPlot}
we plot the angular distribution of the cross section
for $E=170$ [GeV] and $g=3$. As in the $ZZ$ case,
the cross section shows a double-hump structure,
with moderate depletion near the beam
direction and the direction perpendicular to it.

\section{Conclusion}

We have analyzed in this paper various
perturbative mechanisms for the decay
of electroweak gauge bosons into jet events
containing a rapidity gap. At LEP200 energies
the dominant contributions come from the
$\gamma^*\gamma^*$ and $\gamma^*Z$ mechanisms.
To fix an idea, for a gap cut $g=4$,
$\gamma^*Z$ is the dominant contribution with
a cross section $\sigma \sim 0.04$ [pb], and
the $\gamma^*\gamma^*$ mechanism contributes
another $\sigma \sim 0.015$ [pb]. For smaller
gap cuts the cross section is larger. However,
if the gap cut is too small, the background
``fake gap" events from random fluctuation of
final-state hadron fragments can be significant.
For the $\gamma^*\gamma^*$ events the jet fragment
distribution typically is peaked towards the
forward and backward beam direction. For the
$\gamma^* Z$ process one would typically observe
a forward or backward jet system recoiling against
a low-rapidity two-jet system on the opposite side,
hence have a lopsided rapidity gap.
It is also likely to observe the
special cases where the virtual photons
are converted into a vector mesons instead of
jet-pairs.

The $WW$ mechanism may also contribute to the
production of gap events. Our analysis indicates
that the maximum event cross section is reached
$5$ to $10$ [GeV] above the threshold. These
events have a relatively smooth distribution
in the scattering angle. The measurement the
experimental gap event cross section here
will help to test the hypothesis that the color-flow
is confined to the collinear jets.
Finally, the $ZZ$ mechanism has a very small cross section
at LEP200, and we do not expect it to contribute
in practice.

We thank Stanley J. Brodsky, Valery A. Khoze and
Joseph Milana for helpful conversations.
This work is supported by Department of Energy
contract DE-FG02-93ER-40762.

\appendix{FOUR-JET KINEMATICS}

All the production mechanisms in this paper
involve the decay of two gauge bosons into
four final-state quarks. The graphic
representation of the particle momenta
is shown in Fig. \ref{Kinematics}.
We denote the momenta of the incoming
positron and electron by $k_1$ and $k_2$,
respectively,
and we will use $q_1, \bar q_1, q_2, \bar q_2$ to
denote the final-state quarks as well as their
momenta.
The momenta carried by the first and
second gauge bosons are denoted by $P_1$ and $P_2$,
Therefore we have
\begin{eqnarray}
P_1&=&q_1+\bar q_1, \nonumber \\
P_2&=&q_2+\bar q_2.
\end{eqnarray}
That is, the $(q_1 \bar q_1)$ pair is the
decay products of the first gauge boson, and
the $(q_2 \bar q_2)$ pair is the
decay products of the second gauge boson.

At the center-of-mass frame,
the momenta of the four quark jets can be
parametrized in terms of seven kinematic
variables: $M_1^2, M_2^2, x_1, x_2, \theta,
\phi_1, \phi_2$;
where $M_1^2=P_1^2, M_2^2=P_2^2$ are the invariant masses of the
$(q_1 \bar q_1)$ and $(q_2 \bar q_2)$ pairs,
$x_1$ and $x_2$ the longitudinal momentum fractions of
$q_1$ within the $(q_1 \bar q_1)$ pair and
$q_2$ within the $(q_2 \bar q_2)$,
$\theta$ is the gauge-boson scattering angle, and
$\phi_1, \phi_2$ the azimuthal angles of $q_1$ and
$q_2$ with respect to the gauge boson direction.
Notice that $\theta$ coincides with the thrust
axis scattering angle for the $\gamma^*\gamma^*$
and $\gamma^*Z$ mechanisms. However, for
the $ZZ$ and $WW$ mechanisms the thrust angle $\Theta$
defined by the direction of $q_1 + \bar q_2$
is different from $\theta$ due to the crossing of
jet-pairs. The following is the explicit representation
of the initial and final four-vectors:
\begin{eqnarray}
k_1
&=&
\frac{\sqrt{s}}
     {2}
\Bigl[ \ 1; \ \hat{\bf k} \
\Bigr] ,
\nonumber
\\
k_2
&=&
\frac{\sqrt{s}}
     {2}
\Bigl[ \ 1; \ - \hat{\bf k} \
\Bigr] ,
\nonumber
\\
q_1
&=&
\frac{1}
     {2}
\left[ \
   x_1 P_1^+ + \bar x_1 P_1^- ; \
   2 \sqrt{x_1 \bar x_1} M_1 \ \hat{\bf p}_{1\perp}
   + \bigl( x_1 P_1^+ - \bar x_1 P_1^-
     \bigr) \ \hat{\bf P} \
\right] ,
\nonumber
\\
\bar q_1
&=&
\frac{1}
     {2}
\left[ \
   \bar x_1 P_1^+ +  x_1 P_1^- ; \
   - 2 \sqrt{x_1 \bar x_1} M_1 \ \hat{\bf p}_{1\perp}
   + \bigl( \bar x_1 P_1^+ - x_1 P_1^-
     \bigr) \ \hat{\bf P} \
\right] ,
\nonumber
\\
q_2
&=&
\frac{1}
     {2}
\left[ \
   x_2 P_2^+ + \bar x_2 P_2^- ; \
   2 \sqrt{x_2 \bar x_2} M_2 \ \hat{\bf p}_{2\perp}
   - \bigl( x_2 P_2^+ - \bar x_2 P_2^-
     \bigr) \ \hat{\bf P} \
\right] ,
\nonumber
\\
\bar q_2
&=&
\frac{1}
     {2}
\left[ \
   \bar x_2 P_2^+ + x_2 P_2^- ; \
   - 2 \sqrt{x_2 \bar x_2} M_2 \ \hat{\bf p}_{2\perp}
   - \bigl( \bar x_2 P_2^+ - x_2 P_2^-
     \bigr) \ \hat{\bf P} \
\right] .
\end{eqnarray}
It is convenient to choose the $z$ axis along the
direction of the momentum of the first gauge boson
$P_1$, This will facilitate the calculation of
spinor products later. Using the coordinate system shown
in Fig. \ref{Kinematics}, we have
\begin{eqnarray}
\hat{\bf k}
&=&
\cos\theta \hat{\bf z}
-\sin\theta \hat{\bf x},
\nonumber
\\
\hat{\bf p}_{1\perp}
&=&
\cos\phi_1 \hat{\bf x}
+\sin\phi_1 \hat{\bf y},
\nonumber
\\
\hat{\bf p}_{2\perp}
&=&
\cos\phi_2 \hat{\bf x}
+\sin\phi_2 \hat{\bf y},
\nonumber
\\
\hat{\bf P}
&=&
\hat{\bf z} .
\end{eqnarray}
The quantities $x_1, \bar x_1, x_2, \bar x_2, P_1^+,
P_1^-, P_2^+, P_2^-$ satisfy the following constraints
\begin{eqnarray}
x_1 + \bar x_1 &=& 1 , \nonumber \\
x_2 + \bar x_2 &=& 1 , \nonumber \\
P_1^+P_1^- &=& M_1^2 , \nonumber \\
P_2^+P_2^- &=& M_2^2 .
\end{eqnarray}
{}From conservation of energy and momentum, we can
obtain $P_1^+, P_1^-, P_2^+, P_2^-$ explicitly
in terms of $M_1^2, M_2^2$ and $s$
\begin{eqnarray}
P_1^+
&=&
\frac{1}
     {2 \sqrt{s}}
\Bigl[ \
  s + M_1^2 - M_2^2 + \Delta( s, M_1^2, M_2^2 ) \
\Bigr] ,
\nonumber
\\
P_1^-
&=&
\frac{1}
     {2 \sqrt{s}}
\Bigl[ \
  s + M_1^2 - M_2^2 - \Delta( s, M_1^2, M_2^2 ) \
\Bigr] ,
\nonumber
\\
P_2^+
&=&
\frac{1}
     {2 \sqrt{s}}
\Bigl[ \
  s - M_1^2 + M_2^2 + \Delta( s, M_1^2, M_2^2 ) \
\Bigr] ,
\nonumber
\\
P_2^-
&=&
\frac{1}
     {2 \sqrt{s}}
\Bigl[ \
  s - M_1^2 + M_2^2 - \Delta( s, M_1^2, M_2^2 ) \
\Bigr] ,
\end{eqnarray}
where $\Delta( s, M_1^2, M_2^2 )$ is the triangular
function defined in Eq. (\ref{TriangularFunction}).

The spinor products (see Ref. \cite{ManganoParke} for
a review on the helicity method) between the
various  four-vectors can be computed explicitly.
We have
\begin{eqnarray}
\langle k_1 k_2
\rangle
&=&
-\sqrt{s} ,
\nonumber
\\
\langle k_1 q_1
\rangle
&=&
- \Bigl[ \sqrt{s} P_1^+ x_1
  \Bigr]^{1/2} \sin\frac{\theta}{2} \
- \Bigl[ \sqrt{s} P_1^- \bar x_1
  \Bigr]^{1/2} \cos\frac{\theta}{2} \ e^{i \phi_1} ,
\nonumber
\\
\langle k_1 \bar q_1
\rangle
&=&
- \Bigl[ \sqrt{s} P_1^+ \bar x_1
  \Bigr]^{1/2} \sin\frac{\theta}{2} \
+ \Bigl[ \sqrt{s} P_1^-  x_1
  \Bigr]^{1/2} \cos\frac{\theta}{2} \ e^{i \phi_1} ,
\nonumber
\\
\langle k_1 q_2
\rangle
&=&
- \Bigl[ \sqrt{s} P_2^- \bar x_2
  \Bigr]^{1/2} \sin\frac{\theta}{2} \
- \Bigl[ \sqrt{s} P_2^+ x_2
  \Bigr]^{1/2} \cos\frac{\theta}{2} \ e^{i \phi_2} ,
\nonumber
\\
\langle k_1 \bar q_2
\rangle
&=&
- \Bigl[ \sqrt{s} P_2^- x_2
  \Bigr]^{1/2} \sin\frac{\theta}{2} \
+ \Bigl[ \sqrt{s} P_2^+ \bar x_2
  \Bigr]^{1/2} \cos\frac{\theta}{2} \ e^{i \phi_2} ,
\nonumber
\\
\langle k_2 q_1
\rangle
&=&
  \Bigl[ \sqrt{s} P_1^+ x_1
  \Bigr]^{1/2} \cos\frac{\theta}{2} \
- \Bigl[ \sqrt{s} P_1^- \bar x_1
  \Bigr]^{1/2} \sin\frac{\theta}{2} \ e^{i \phi_1} ,
\nonumber
\\
\langle k_2 \bar q_1
\rangle
&=&
  \Bigl[ \sqrt{s} P_1^+ \bar x_1
  \Bigr]^{1/2} \cos\frac{\theta}{2} \
+ \Bigl[ \sqrt{s} P_1^-  x_1
  \Bigr]^{1/2} \sin\frac{\theta}{2} \ e^{i \phi_1} ,
\nonumber
\\
\langle k_2 q_2
\rangle
&=&
  \Bigl[ \sqrt{s} P_2^- \bar x_2
  \Bigr]^{1/2} \cos\frac{\theta}{2} \
- \Bigl[ \sqrt{s} P_2^+ x_2
  \Bigr]^{1/2} \sin\frac{\theta}{2} \ e^{i \phi_2} ,
\nonumber
\\
\langle k_2 \bar q_2
\rangle
&=&
  \Bigl[ \sqrt{s} P_2^- x_2
  \Bigr]^{1/2} \cos\frac{\theta}{2} \
+ \Bigl[ \sqrt{s} P_2^+ \bar x_2
  \Bigr]^{1/2} \sin\frac{\theta}{2} \ e^{i \phi_2} ,
\nonumber
\\
\langle q_1 \bar q_1
\rangle
&=&
M_1 \ e^{i \phi_1} ,
\nonumber
\\
\langle q_2 \bar q_2
\rangle
&=&
M_2 \ e^{i \phi_2} ,
\nonumber
\\
\langle q_1 q_2
\rangle
&=&
  \Bigl[ \bar x_1 \bar x_2 P_1^- P_2^-
  \Bigr]^{1/2} \ e^{i \phi_1}
- \Bigl[ x_1 x_2 P_1^+ P_2^+
  \Bigr]^{1/2} \ e^{i \phi_2} ,
\nonumber
\\
\langle q_1 \bar q_2
\rangle
&=&
  \Bigl[ \bar x_1 x_2 P_1^- P_2^-
  \Bigr]^{1/2} \ e^{i \phi_1}
+ \Bigl[ x_1 \bar x_2 P_1^+ P_2^+
  \Bigr]^{1/2} \ e^{i \phi_2} ,
\nonumber
\\
\langle \bar q_1 q_2
\rangle
&=&
- \Bigl[ x_1 \bar x_2 P_1^- P_2^-
  \Bigr]^{1/2} \ e^{i \phi_1}
- \Bigl[ \bar x_1 x_2 P_1^+ P_2^+
  \Bigr]^{1/2} \ e^{i \phi_2} ,
\nonumber
\\
\langle \bar q_1 \bar q_2
\rangle
&=&
- \Bigl[ x_1 x_2 P_1^- P_2^-
  \Bigr]^{1/2} \ e^{i \phi_1}
- \Bigl[ \bar x_1 \bar x_2 P_1^+ P_2^+
  \Bigr]^{1/2} \ e^{i \phi_2} .
\end{eqnarray}
All other spinor products can be obtained by
the antisymmetry operation or by
helicity conjugation:
\begin{eqnarray}
\langle a b \rangle
&=& - \langle b a \rangle
\nonumber
\\
 \ [ a b ]
&=&
 -[ b a ] = \langle b a \rangle ^*
\end{eqnarray}

\figure{\label{QCDMechanism}
Perturbative QCD mechanisms for generating
rapidity-gap events at the $Z$ peak. The dashed lines indicate
that the produced partons are in color-singlet
state. (a) Two final-state quark-antiquark pairs.
(b) A quark-antiquark jet pair and a two-gluon
jet pair.
}

\figure{\label{QEDMechanism}
A QED mechanism for generating rapidity-gap
events at the $Z$ peak. The dashed lines indicate
that the produced partons are in color-singlet
state.
}

\figure{\label{GammaGammaMechanism}
The $e^+e^- \to \gamma^*\gamma^* \to
q\bar qq\bar q$ mechanism for generating
gap events.
}

\figure{\label{GammaGammaLego}
A typical Lego plot for a gap event from
the $\gamma^*\gamma^*$ mechanism. The
physically observed gap is expected
to be $g_{\rm eff} \simeq g-1.4$ due to
the spreading of hadron fragments around
each jet.
}

\figure{\label{GammaGammaGapEnergyPlot}
Gap event cross section for the $\gamma^*\gamma^*$
mechanism, as function of the gap-cut $g$ and
for two different values of energies.
}

\figure{\label{GammaGammaEnergyGapPlot}
Gap event cross section for the $\gamma^*\gamma^*$
mechanism, as function of the energy and
for different values of the gap-cut $g$.
}

\figure{\label{ZPeakGammaGammaCrossSection}
The $\gamma^*\gamma^*$ contribution to the gap-event
cross section at the $Z$-peak (solid line). Also plotted
are the contributions from the QCD (dashed line) and
QED (dotted line) mechanisms (see Fig. \ref{QCDMechanism},
Fig. \ref{QEDMechanism} and also Ref. \cite{LuBrodskyKhoze}.)
}

\figure{\label{GammaZMechanism}
The $e^+e^- \to \gamma^*Z \to
q\bar qq\bar q$ mechanism for generating
gap events at LEP200.
}

\figure{\label{GammaZLego}
A typical Lego plot for a gap event from
the $\gamma^*Z$ mechanism. Notice that
the rapidity gap is shifted towards the
virtual photon side.
}

\figure{\label{GammaZGapEnergyPlot}
Gap event cross section for the $\gamma^*Z$
mechanism, as function of the gap-cut $g$ and
for two different values of energies.
}

\figure{\label{GammaZEnergyGapPlot}
Gap event cross section for the $\gamma^*Z$
mechanism, as function of the energy and
for different values of the gap-cut $g$.
}

\figure{\label{ZZMechanism}
The $e^+e^- \to Z Z \to
q\bar qq\bar q$ mechanism for generating
gap events at LEP200. Notice the
exchange of quarks and antiquarks from
different gauge bosons
to form color-singlet jet-pairs
}

\figure{\label{ZZEnergyGapPlot}
Gap event cross section for the $ZZ$
mechanism, as function of the energy and
for different values of the gap-cut $g$.
}

\figure{\label{ZZAngularDistributionPlot}
Angular distribution of gap event cross section
for the $ZZ$ mechanism, where $\Theta$ is
the thrust axis scattering angle.
}

\figure{\label{WWMechanism}
The $e^+e^- \to W W \to
q\bar qq\bar q$ mechanism for generating
gap events at LEP200. Notice the
exchange of quarks and antiquarks from
different gauge bosons
to form color-singlet jet-pairs
}

\figure{\label{WWEnergyGapPlot}
Gap event cross section for the $WW$
mechanism, as function of the energy and
for different values of the gap-cut $g$.
}

\figure{\label{WWAngularDistributionPlot}
Angular distribution of gap event cross section
for the $WW$ mechanism, where $\Theta$ is
the thrust axis scattering angle.
}

\figure{\label{Kinematics}
Initial and final momenta involved in
$e^+e^- \to q\bar qq\bar q$.
}

\end{document}